\documentclass[10pt,a4paper]{article}
\usepackage{graphicx}
\date{}
\usepackage[latin1]{inputenc}
\usepackage{amsmath}
\usepackage{amsfonts}
\usepackage{amssymb}
\begin{document}
\title{\textbf{Magnetic monopole and string excitations in two-dimensional spin
ice}}
\author{{L. A. M\'{o}l}$^2$, {R. L.  Silva} $^{1}$, {R. C. Silva}$^{1}$, {A. R. Pereira}$^{1}$ , {W.A. Moura-Melo}$^{1}$ ,
B.V. Costa$^{2}$ \\ \\
\small $^{1}$ \it
Departamento de F\'{\i}sica, Universidade Federal de Vi\c{c}osa\\
\small \it 36570-000, Vi\c{c}osa, Minas
Gerais, Brazil\\
\small $^{2}$ \it Departamento de F\'{\i}sica, ICEX, Universidade
Federal de Minas Gerais\\ \small \it  Caixa Postal 702, 30123-970,
Belo Horizonte, Minas Gerais, Brazil} \maketitle
\begin{center}
\textbf{Abstract}
\end{center}

\indent We study the magnetic excitations of a square lattice
spin-ice recently produced in an artificial form, as an array of
nanoscale magnets. Our analysis, based upon the dipolar
interaction between the nanomagnetic islands, correctly reproduces
the ground-state observed experimentally. In addition, we find
magnetic monopole-like excitations effectively interacting by
means of the usual Coulombic plus a linear confining potential,
the latter being related to a string-like excitation binding the
monopoles pairs, what indicates that the fractionalization of
magnetic dipoles may not be so easy in two dimensions. These
findings contrast this material with the three-dimensional
analogue, where such monopoles experience only the Coulombic
interaction. We discuss, however, two entropic effects that affect
the monopole interactions: firstly, the string configurational
entropy may loose the string tension and then, free magnetic
monopoles should also be found in lower dimensional spin ices;
secondly, in contrast to the string configurational entropy, an
entropically driven Coulomb force, which increases with
temperature, has the opposite effect of confining the magnetic
defects.
Corresponding author: A. R. Pereira\\
e-mail: apereira@ufv.br\\
Tel.: +55-31-3899-2988.\\
Fax: +55-31-3899-2483.\\


\section{Introduction}

 Geometrical frustration among spins in magnetic materials
can lead to a variety of cooperative phases such as spin glass,
spin liquid and spin ice behaving like glass, liquid and ice in
nature. The description and understanding of such states are
becoming increasingly important not only in condensed matter but
also, in other branches like field theories. In a crystal at low
temperature excitations above the ground state often behave like
elementary particles carrying a quantized amount of energy,
momentum, electric charge and spin. Several of these objects arise
as a result of the collective behavior of many particles in a
material which is most effectively described in terms of the
fractions of the original particles. The emergence of these
excitations is an example of the phenomenon known as
``fractionalization". This occurrence is often tied to topological
defects \cite {Goldstone81} and is common in one-dimensional
systems (polyacetylene, nanotubes, etc). Higher dimensional
fracionalization is more difficult to be found. In two spatial
dimensions the only confirmed case is the involvement of
quasi-particles with one-third of an electron's charge in the
fractional quantum Hall effect in strong magnetic fields. Among
several suggestions \cite{Nussinov07}, there is also the proposal
that the merons forming a skyrmion in two-dimensional ($2d$)
Heisenberg antiferromagnets are spinons and therefore, they are
neutral spin-half excitations \cite{Baskaran03,Antonio07}. More
recently, examples of fractionalization in three-dimensional
systems were provided in spin ice materials
\cite{Nussinov07,Castelnovo08}. Particularly, the authors of
Ref.\cite{Castelnovo08} have shown how the famous magnetic
monopole may emerge in these materials. Despite some exciting
suggestions for its existence from the realms of quantum
mechanics, a single magnetic pole remains elusive after decades of
searching in particle accelerators and cosmic rays. Now,
Castelnovo \emph{et al.} \cite{Castelnovo08} indicated an
unexpected but, perhaps, better place to look. Under certain
conditions, spin ice magnets behave like a gas of free magnetic
poles. There is even a phase transition at which a thin vapor of
these poles condenses into a dense liquid. An experimentally
measurable signature of monopole dynamics on a diamond lattice in
the grand canonical ensemble was presented in
Ref.\cite{Joubert09}. The existence of these excitations in a
condensed matter system is exciting in itself. Our aim in this
paper is to study spin ice materials, but in two spatial
dimensions. Such structures have been artificially produced in a
geometrically frustrated lattice of nanoscale ferromagnetic
islands \cite{Wang06,Remhof08,Ke08}. Here, we examine the
excitations (``magnetic monopoles") and how they interact in this
$2d$ system.

\begin{figure}
\centering
\includegraphics[angle=0.0,width=10cm]{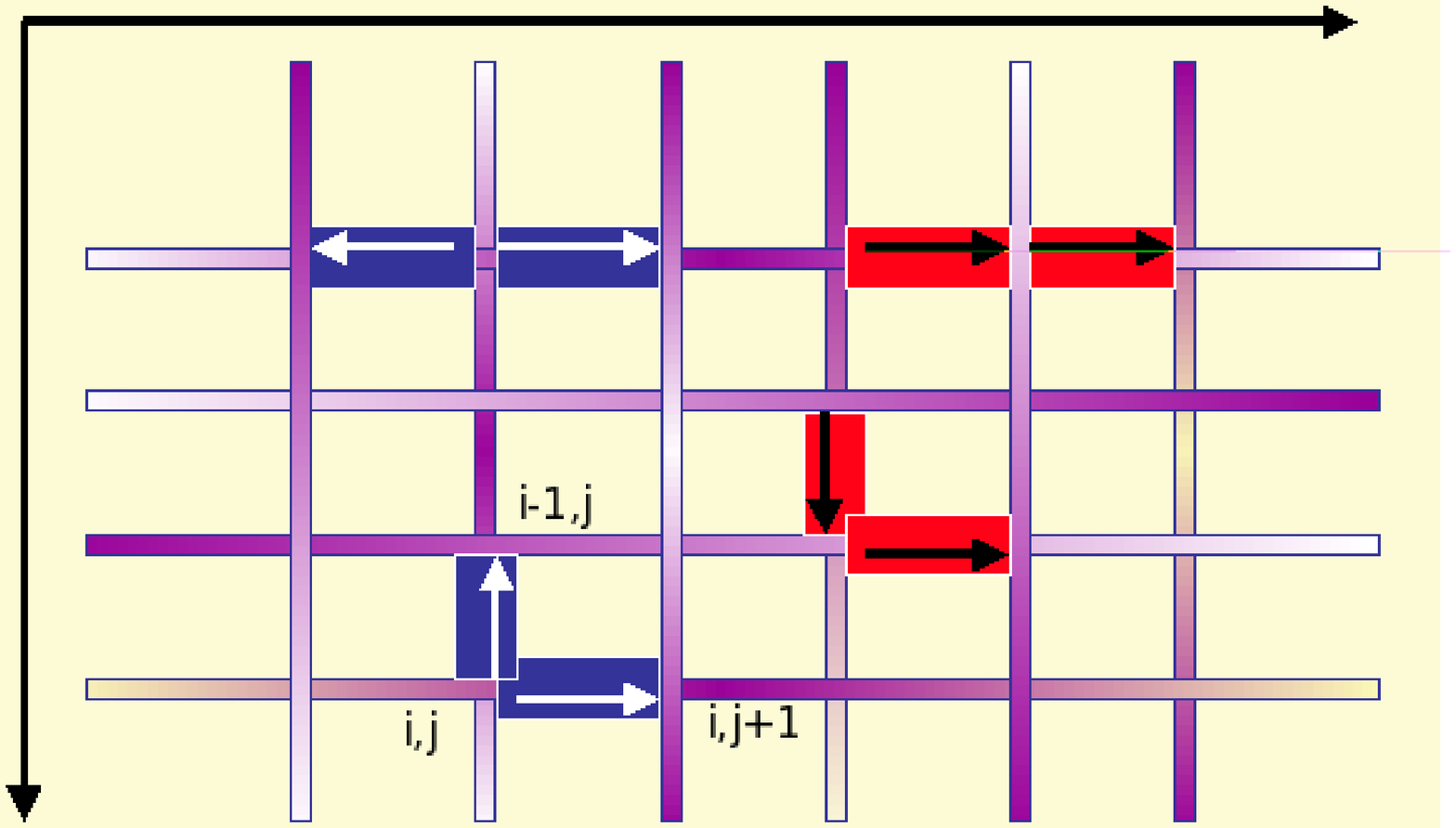}
\caption{} \label{Spinice}
\end{figure}

Three-dimensional spin-ice materials have the pyrochlore structure
in which magnetic rare-earth ions form a lattice of corner-sharing
tetrahedra. To minimize the spin-spin interaction energy, the ice
rules are manifested: two spins point inward and two spins point
outward on each tetrahedron. A similar system was built in two
dimensions with elongated permalloy nanoparticles. This artificial
material consists of elongated magnetic nano-islands distributed in
a 2d square lattice. The longest axis of the islands alternate its
orientation pointing in the direction of the two principal axis of
the lattice \cite{Wang06}. The magnetocrystalline anisotropy of
permalloy is effectively zero, so that the shape anisotropy of each
island forces its magnetic moment to align along the largest axis
thus, making the islands effectively Ising-like. The intrinsic
frustration on this lattice is similar to that in the spin ice model
and can be best seen by considering a vertex at which four islands
meet. A pair of moments on a vertex can be aligned either to
maximize or to minimize the dipole interaction energy of the pair.
As shown in Ref.\cite{Wang06}, it is energetically favorable when
the moments of a pair of islands are align so that one is pointing
into the center of the vertex and the other is pointing out (red
islands in Fig.\ \ref{Spinice}) while it is energetically
unfavorable when both moments are pointing inward or both are
pointing outward (blue islands in Fig.\ \ref{Spinice}). This
artificial system exhibits short-range order and ice-like
correlations on the lattice, which is precisely analogous to the
behavior of the spin ice materials. However, it should be stressed
that the fundamental interaction among the islands is the long range
dipole-dipole force, once the short-ranged exchange is negligible in
this case, where the islands are spaced by around $320\,{\rm nm}$,
much greater than Permalloy exchange length, around $5-7\,{\rm nm}$.
Here, we consider an arrangement alike that experimentally
investigated in Ref. \cite{Wang06}. In our scheme the magnetic
moment (``spin") of the island is replaced by a point dipole at its
center. To do this, in each site $(x_{i},y_{i})$ of a square lattice
two spin variables are defined: $\vec{S}_{h(i)}$ with components
$S_{x}=\pm 1$, $S_{y}=0$ located at $\vec{r}_{h}=(x_{i}+1/2,y_{i})$,
and $\vec{S}_{v(i)}$ with components $S_{x}=0$, $S_{y}=\pm 1$ at
$\vec{r}_{v}=(x_{i},y_{i}+1/2)$. Therefore, in a lattice of volume
$L^{2}=l^{2}a^{2}$ ($a$ is the lattice spacing) one gets $2\times
l^{2}$ spins (see Fig.\ \ref{groundstate}). Representing the spins
of the islands by $\vec{S}_{i}$, which can assume either
$\vec{S}_{h(i)}$ or $\vec{S}_{v(i)}$, then the $2d$ spin ice is
described by the following Hamiltonian
\begin{eqnarray}\label{HamiltonianSI}
H_{SI} &=& Da^{3} \sum_{i\neq j}\left[\frac{\vec{S}_{i}\cdot
\vec{S}_{j}}{r_{ij}^{3}} - \frac{3 (\vec{S}_{i}\cdot
\vec{r}_{ij})(\vec{S}_{j}\cdot \vec{r}_{ij})}{r_{ij}^{5}}\right],
\end{eqnarray}\\
where $D=\mu_{0}\mu^{2}/4\pi a^{3}$ is the coupling constant of
the dipolar interaction. The sum is either over all
$l^2(2l^{2}-1)$ pairs of spins in the lattice for the case with
open boundary conditions (OBC) or over all spins and their images
for the case with periodic boundary conditions (PBC). We study
these two possibilities; OBC is more related to the artificial
spin ice fabricated in Ref. \cite{Wang06}, while using PBC we
minimize the border effects. In the system with PBC the Ewald
summation \cite{Weis03,Wang01} is used.

\section{The model and results}

To start, we consider the ground states obtained from Hamiltonian
(\ref{HamiltonianSI}) describing the $2d$ spin ice. To do this we
use a simulated annealing process \cite{Landau05}, which is a
Monte Carlo calculation where the temperature is slightly reduced
in each step of the process in order to drive the system to the
global minimum. Our Monte Carlo scheme consist of a simple
Metropolis algorithm \cite{Landau05}. In each Monte Carlo step
(MCS) we attempt to flip all spins in the lattice sequentially or
randomly which gives the same results. Several tests for systems
with different sizes $L$ ($6a \leq L\leq 80a$) were studied. In
each simulation $10\times l^2$ Monte Carlo steps were done at each
temperature starting at $T=3.0$ and decreasing the temperature in
steps $\Delta T=0.2$ until $T=0.2$ (the temperature is measured in
units of $D/k_B$). We observed that for $T<0.4$ the system
freezes, in the sense that all trial moves are rejected. The final
configuration (ground state) was found to be twofold degenerate
(see part (a) of Fig. \ \ref{groundstate} for a lattice with
$L=6a$). If we consider the vorticity in each plaquette, assigning
a variable $\sigma=+1$ and $-1$ to clockwise and anticlockwise
vorticities respectively, the ground state looks like a
checkerboard, with an antiferromagnetic arrangement of the
$\sigma$ variable. Note that the ground state clearly obeys the
ice rule. We remark that it is impossible to minimize all
dipole-dipole interactions. Actually, on each vertex there are six
pairs of dipoles and only four of them can simultaneously minimize
the energy. It is important to mention that, although there are
other possible configurations that also obey ice rules, these are
not the ground state. Indeed, the state shown on the right side of
Fig.\ \ref{groundstate} has energy about four times larger than
that of the ground state. The difference between these two states
is related to the distinct topologies for the configurations of
the four moments (see Fig.\ \ref{Topology}). It was experimentally
shown in Ref. \cite{Wang06} that, while the topologies of types
(a) and (b) obey the ice rule, the case (a) has smaller energy
than case (b). Our theoretical calculations confirm this fact. The
same ground state was also reported in Refs.\cite{Remhof08,Ke08}.
We would like to remark that although this is the ground state,
its thermal equilibration in experiments seems to be very
difficult \cite{Remhof08,Ke08,Moller06}.

Once the system is naturally frustrated, in the two-in/two-out
configuration, the effective magnetic charge $Q_{M}^{i,j}$ (number
of spins pointing inward minus the number of spins pointing outward
on each vertex $(i,j)$) is zero everywhere. The most elementary
excited state involves inverting a single spin to generate localized
``dipole magnetic charges", which implies in a ``vortex-pair
annihilation". Such an inversion corresponds to two adjacent sites
with net magnetic charge $Q_{M}^{i,j}=\pm 1$, which is alike a
nearest-neighbor monopole-antimonopole pair. In principle, such
``monopoles" can be separated from one another without violations of
local neutrality by flipping a chain of adjacent spins. One can
easily see that in this process a ``string" of spins pointing from
the positive to the negative charge is created (see Fig.\
\ref{cena10}). The presence of a string-like excitation joining
these poles is evidenced by an extra energy cost behaving as $bX$,
where $X$ is the length of the string and $b>0$ is the effective
string tension, as below.
%
In order to establish a link between the monopole-antimonopole
distance $R$ and the string length $X$ we choose two basic string
shapes to move the charges as shown in Fig.\ \ref{cena10}. Of
course, the shortest strings will be formed around the straight line
joining the monopoles and, therefore, we choose two different ways
in which they may be created as the charges are separated (see Fig.\
\ref{cena10}). Firstly, using the string shape $1$ and starting in
the ground state we choose an arbitrary site and then the gray spins
in Fig.\ \ref{cena10} are flipped, so creating a
monopole-antimonopole separated by $R=2a$. In sequence, the spins
marked in blue are flipped and the separation distance becomes
$R=4a$ and so on. In this case $X=4R/2$. Note that the string surges
in the system because in the separation process, the topology is
locally modified, although still keeping the ice rule; in the region
between the two poles, the topology of type (b), which has larger
energy than that of type (a), prevails. Being essentially localized
along the line joining the monopoles this additional amount of
energy increases as the distance between the magnetic charges
increases, justifying the $bX$ term.

The potential $V(r)$ (the energy of the excited configuration
minus the energy of the ground state) as a function of $r=R/a$ can
be obtained by simple evaluation of the energy of each
configuration. It is shown in the inset of Fig.\ \ref{potential}
for the string shape presented in part (1) of Fig.\ \ref{cena10}.
The behavior is apparently linear but the function $f_{q}(R)=q/R
+b'R +c$, with $q\approx -0.00122Da, b=b'/2 \approx 0.00305 D/a,
c\approx 0.00734 D$, fits better the data than the purely linear
possibility $g(R)=\alpha R + \beta$, with $\alpha\approx 0.00611
D/a, \beta\approx 0.00702D$. This difference becomes clearer when
we analyze the $\chi^2/Dof$ which is equal to $1.04 \times
10^{-8}$ for the linear fitting and $4.5\times 10^{-13}$ for
$f_{q}(R)$. Also, in Fig.\ \ref{potential} we draw a baseline of
the potential using the linear fit. One can clearly see that
$f_{q}(R)$ describes better the data and, therefore, $V(r)\approx
f_{q}(R)$. The same method was repeated using the string shape
$2$. In this case, the charges are separated diagonally and
$X=2R/\sqrt{2}$. The results are qualitatively the same and the
values of the constants are: $q\approx -0.00125Da,
b=b'/\sqrt{2}\approx 0.00317 D/a, c\approx 0.00724 D$. Note that
the quantitative changes are small. The results are also
qualitatively the same if PBC are used instead of OBC.
Furthermore, quantitative differences between PBC and OBC
calculations are smaller than $1\%$ for constants $b$ and $c$,
while it is smaller than $9\%$ for $q$. The larger difference for
constant $q$ can be understood if one remembers that the use of
PBC will imply that the charges interact also with their images.

Our calculations yield the total energy cost of a
monopole-antimonopole pair, separated by $R$, as the sum of the
usual Coulombic-type term, $q/R$ ($q<0$ is a constant), and an extra
contribution behaving like $bX$, brought about from the string-like
excitations that bind the monopoles, so that, $V(R)=q/R+bX(R)+c$
($X(R)$ is the string length, while $c$ is a constant associated to
the monopole pair creation). Until now we have only considered the
shortest strings connecting two poles. However, many dipole strings
of arbitrary shape and size can be identified that connect a given
pair of monopoles. The associated energy cost increases with $X$ and
diverges with the length of the string. So, at first sight, the
monopoles should be confined in the artificial material. As we will
argue later, it is possible that the string tension vanishes at a
critical temperature proportional to $b$ and hence, free magnetic
monopoles may also be found in the $2d$ system.

For concreteness, the magnetic charge may be easily estimated if we
take into account experimental values of some parameters.
Considering the usual expression for the Coulombic interaction (in
MKS units) $-\mu_{0}Q_{M}^{2}/4\pi R$, we get, $\mid q
\mid=\mu_{0}Q_{M}^{2}/4\pi$, or $Q_{M}\approx \pm \sqrt{4\pi \mid q
\mid/\mu_{0}}\sim \pm 0.035 \mu/a$. Now, using data from Ref.
~\cite{Wang06} (such as $a\sim 320nm$ and $\mu\sim 2.79 \times
10^{-16} JT^{-1}$), the fundamental magnetic charge of an excitation
in the array of ferromagnetic nano-islands reads $Q_{M}\approx 3
\times 10^{-11} Cm/s$, which is about $6\times 10^{3}$ times smaller
than the fundamental charge of the Dirac monopole ($Q_{D}=2\pi
\hbar/\mu_{0}e$). Such a charge can even be tuned continuously by
changing the lattice spacing.

\section{Discussion}

Before concluding, it is important to analyze the behavior of the
string tension as some parameters are varied in the system. The
string tension for the artificial system built in Ref.
\cite{Wang06} is approximately given by $b\approx 2.26 \times
10^{-15}J/m \approx 4.5 \times 10^{-3} eV/a$. Therefore, it is
necessary a relatively large amount of energy (about $10^{-3} eV$)
to separate the ``two-dimensional monopoles" by one lattice
spacing, regardless of how far apart they are. Consequently, at
low temperature, there is insufficient thermal energy to create
long strings, and so the ``monopoles" would be bound together
tightly in pairs. The string tension can be artificially reduced
by increasing the parameter $a$ ($b \propto 1/a$). However, it has
also the effect of decreasing the magnetic charge since $Q_{M}$ is
proportional to $1/a$. A way to reduce $b$, without affecting
$Q_{M}$, is increasing the temperature. By using the random walk
argument, one can see that the many possible ways of connecting a
pair of monopoles with a string give rise to a string
configurational entropy proportional to $R$. Then, as the
temperature increases, the string tension should decrease like
$b-\epsilon k_{B}T$, with $\epsilon=O(a^{-1})$. It means that the
string may loose its tension by entropic effect and, therefore, it
should vanish at some critical temperature $k_{B}T_{c}$, of the
order of $ba\sim 4.5 \times 10^{-3}eV$. Another important point in
this discussion is that as the temperature increases the monopoles
density also increases. Indeed, if the pair creation energy is of
the order of $E_{c}= V(a) \sim 1.3 \times 10^{-2} eV$, one expects
that, for temperatures above this value, the description in terms
of monopoles itself could break down (in fact, the Boltzmann
factor $\exp(-\beta E_{c})$ would increase considerably for
$k_{B}T>>E_{c}$). Then, a possible deconfined phase would live
between the melting temperature of the ordered and the dense
monopole phases. A comparison between $ba$ and $E_{c}$ suggests
that a temperature window between the confined and deconfined
phases could be perfectly plausible in the range $4.5 \times
10^{-3}eV < k_{B}T < 1.3 \times 10^{-2}eV$. Our expectation is
that the window is still greater (starting at a much lower
temperature) since the argument based on the balance of energy
versus entropy may overestimate the critical temperature (for
instance, the Berezinskii-Kosterlitz-Thouless critical temperature
estimated by this argument for the planar rotator model is much
higher than the correct value obtained by Monte-Carlo simulations,
which is $T_{BKT}\simeq 0.89 J$ \cite{Tobochnik79}, where $J$ is
the coupling constant of the model). Once the deconfinement
realizes, the question of technological applications of this
system is relevant. For instance, learning how to move the
magnetic monopoles around would be of importance towards
technologies involving magnetic analogous of electric circuits.

However, there is another entropic effect, discussed in previous
works of purely ice-rule problem and related short-range problems
\cite{Kondev98, Fisher63, Moore99,Henley05} for strictly $2d$
systems, that may change the scenario of free monopoles. That is
the entropic interactions between monopoles due to the underlying
spin configuration. Really, two monopoles should be attracted
because there are more ways to arrange the surrounding dipoles in
the lattice when they are close together. These entropic
interactions, in a strictly $2d$ system, results in a $2d$
effective Coulomb attraction like $\ln R$, between oppositely
charged monopoles, whose strength vanishes proportionally to $T$,
at low temperatures (of course, in three-dimensional materials,
such entropic effect should result in a $1/R$ attraction). In our
case, this logarithmic interaction could be present in addition to
the three-dimensional ($3d$) Coulombic, $q/R$, and linear, $bR$,
interactions discussed in this work. Thus, at temperature high
enough to destroy the string tension, this entropically driven
$2d$ Coulomb force would become crucial for keeping the
monopole-antimonopole pairs bounded, in such a way that no free
monopole phase would occur at any temperature. Nevertheless, how
these monopoles precisely experience such an effect in their local
dynamics is somewhat mysterious and should be investigated in more
details. [We should recall that our calculations to obtain the
interaction potential between monopoles have been performed at
zero temperature and, consequently, this entropic contribution
could not be directly (or even indirectly) present in $V(R)$]. The
precise effect of the temperature on $V(R)$ is under investigation
and will be communicated elsewhere. Here, it should be remarked
that the present monopoles are not actually two-dimensional
objects: their physical interaction is given by the usual
three-dimensional Coulomb force, which means that they should
affect magnetic test particles placed at relatively large
distances along the direction perpendicular to the plane of
islands (we remember that in a strictly $2d$ space, the magnetic
field should be a pseudoscalar field. In addition, a genuine $2d$
``monopole'', as a counterpart of the Dirac pole, appears to be
not magnetic charge; it rather looks like an exotic electric
charge, giving rise to a rotational electric field, instead a
radial-like, as usual charges do. For details, see
Refs.\cite{Teitelboim,Pisarski,PRD2001}). The dipoles forming the
$2d$ lattice are genuinely $3d$ objects and their long-range
dipolar interaction propagates in the $3d$ space, see Hamiltonian
(\ref{HamiltonianSI}). It is an important difference of this
system when compared to the strictly $2d$ models such as vertex
models and others. Now, it is also important to say that, such
entropic interaction will not be accompanied by a magnetic field,
it will not renormalize the monopole charge and it will not be
felt by a stationary magnetic test particle \cite{Castelnovo08}.
Therefore, all calculations concerning the energy scales involved
in the physical interactions between the defects will not be
altered. In addition, it seems that this force has not been
measured directly, yet. The peculiarities between strictly $2d$
models and the system studied here have not been considered
previously. Thus, it is not completely clear if and how the
entropically driven $2d$ Coulomb force acts in the spin ice with
an inherent $3d$ spatial behavior.

\begin{figure}
\centering
\includegraphics[angle=0.0,width=8.0cm]{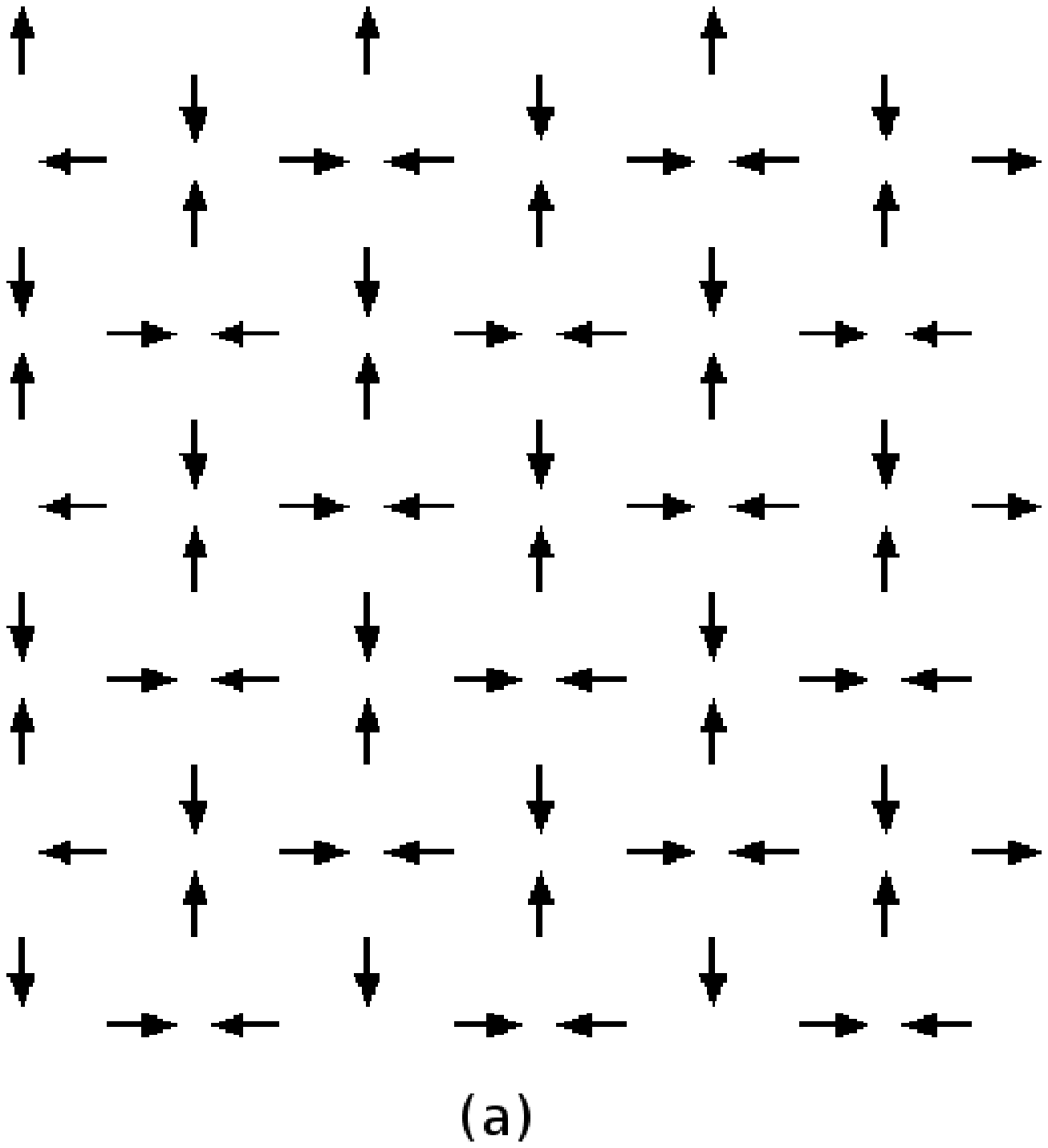}
\includegraphics[angle=0.0,width=8.0cm]{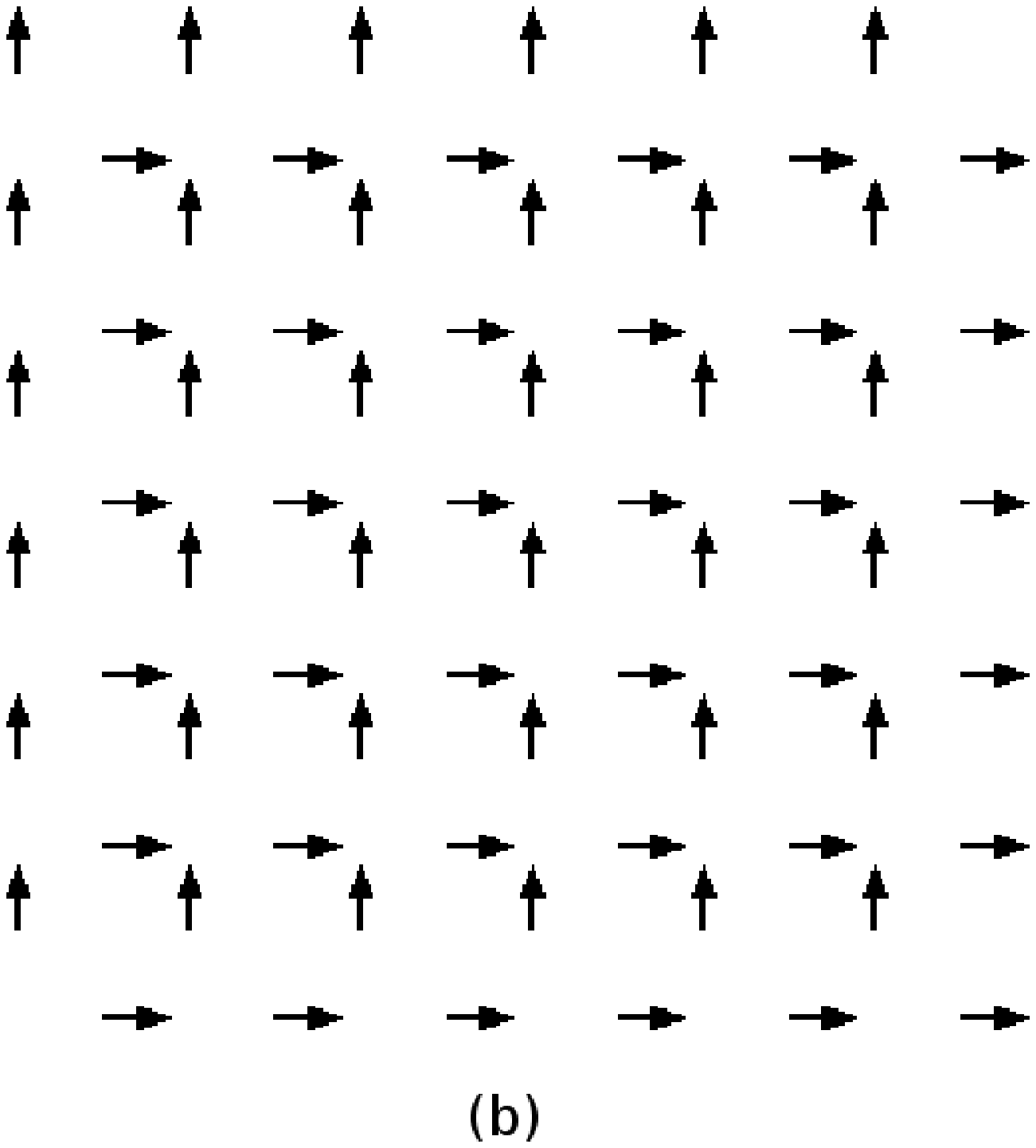}
\caption{} \label{groundstate}
\end{figure}

\begin{figure}
\centering
\includegraphics[angle=0.0,width=15cm]{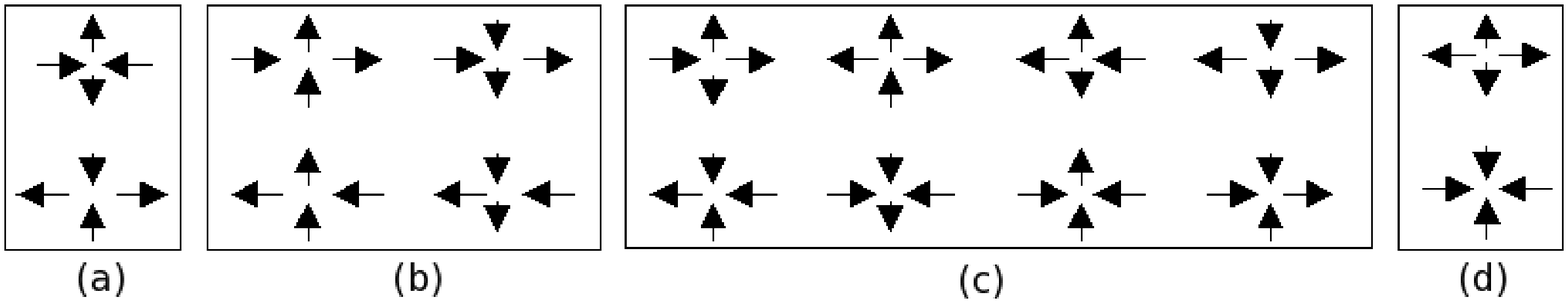}
\caption{} \label{Topology}
\end{figure}

\begin{figure}
\centering
\includegraphics[angle=0.0,width=8.0cm]{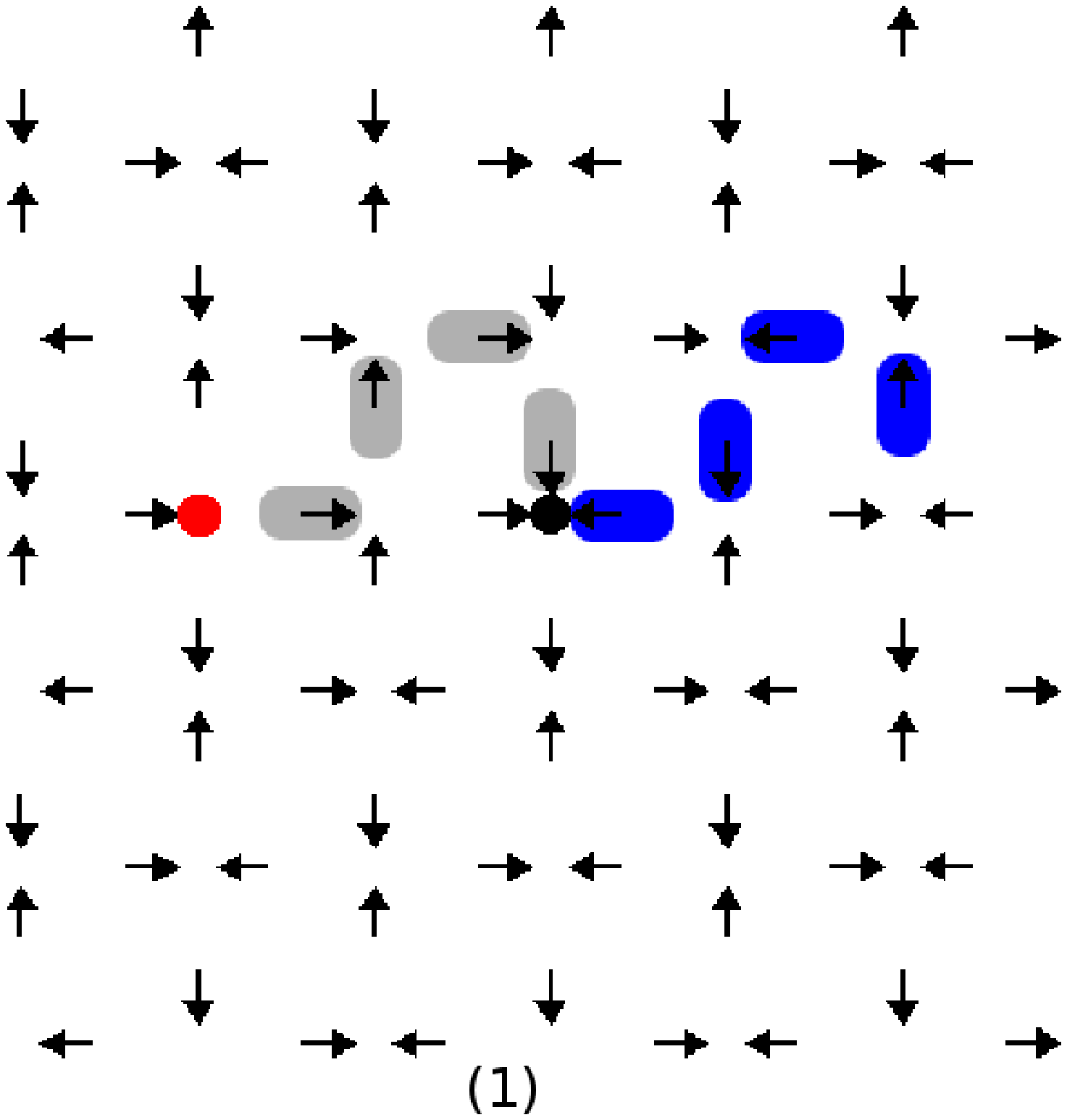}
\includegraphics[angle=0.0,width=8.0cm]{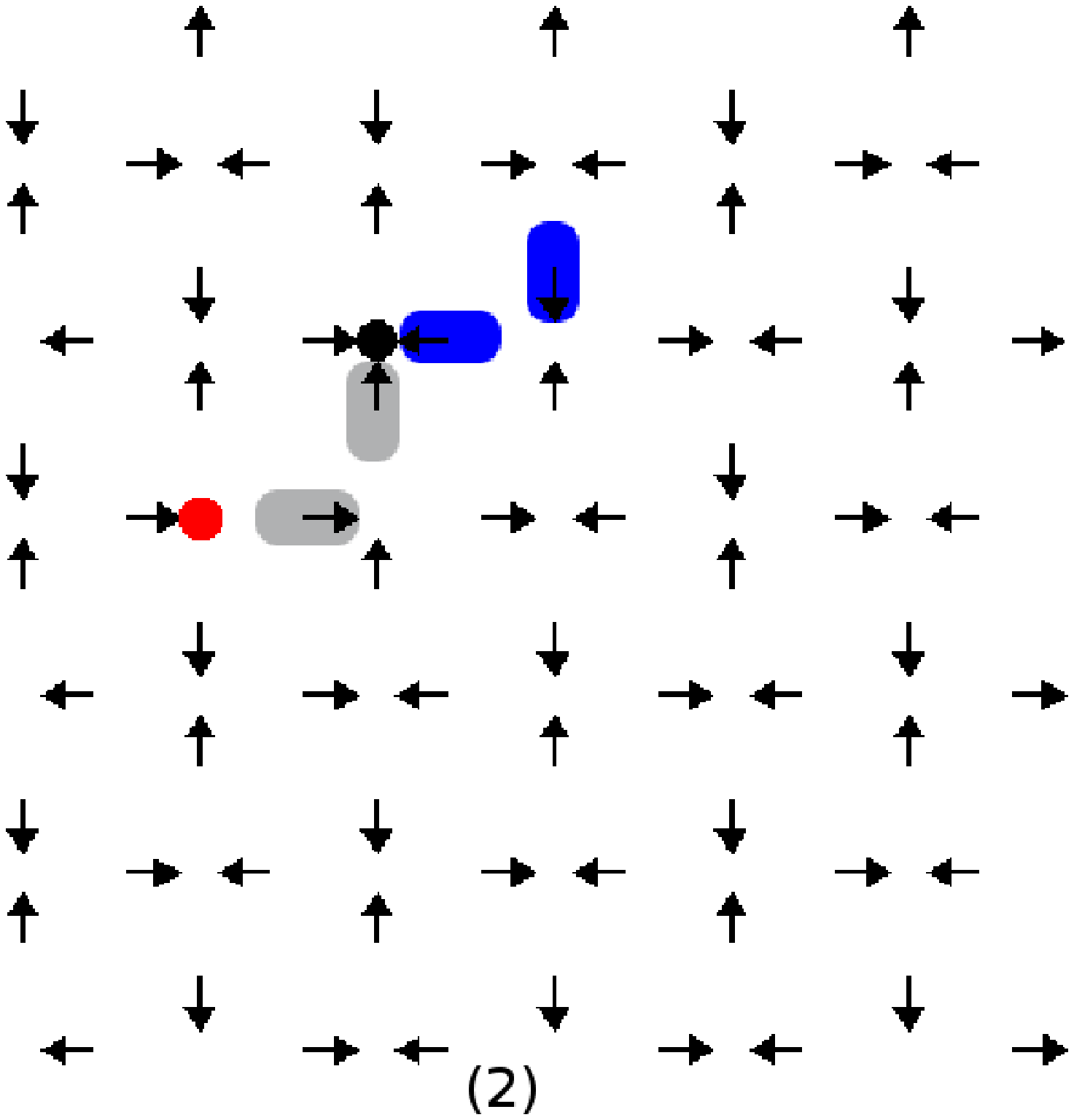}
\caption{} \label{cena10}
\end{figure}

Finally, we should remark that the above scenario involving the
monopole physical interactions may be drastically changed if one
considers these excitations in the configuration (b) of Fig.\
\ref{groundstate}. As experimentally shown in Ref. \cite{Remhof08},
this metastable state is a very real possibility when magnetic
fields are applied. In this case only the topology of type (b) of
Fig.\ \ref{Topology} is present in the separation process. Further
investigation is demanded for shedding extra light on this subjetc.

\vskip .5cm \centerline{\large\bf Acknowledgements} \vskip .3cm The
authors thank CNPq, FAPEMIG and CAPES (Brazilian agencies) for
financial support.
\\
\vskip 1cm


\newpage

Figure Captions


Figure 1. The two-dimensional square lattice studied in this work.
Only a few islands are shown. The arrows inside the islands
represent the local dipole moments ($\vec{S}_{h(i)}$ or
$\vec{S}_{v(i)}$).

Figure 2. (a) Configuration of the ground-state obtained for $L=6a$,
in exact agreement with that experimentally observed. Note that the
ice rules are manifested at each vertex. This is the case in which
the topology demands the minimum energy (see Fig. (\
\ref{Topology})). (b) Another configuration also respecting ice
rules, but displaying a topology which costs more energy.

Figure 3. The 4 distinct topologies and the 16 possible magnetic
moment configurations on a vertex of 4 islands. Although
configurations (a) and (b) obey the ice rule, the topology of (a) is
more energetically favorable than that of (b). Hamiltonian
(\ref{HamiltonianSI}) correctly yields to the true ground-state
based on topology (a), without further assumptions. Topologies (c)
and (d) do not obey the ice rule. Particularly, (c) implies in a
monopole with charge $Q_{M}$.

Figure 4. The two basic shortest strings used in the separation
process of the magnetic charges: pictures (1) and (2) exhibit
strings $1$ and $2$ respectively. The left circle (red) is the
positive charge (north pole) while the right circle (black) is the
negative (south pole).

Figure 5.Inset: the interaction potential between two magnetic
charges (with opposite signs) as a function of $r=R/a$. The baseline
of $V(r)$ is also plotted: the curves are obtained by fitting the
data to $\alpha R + \beta$ and $q/R+b'X +c $ minus $\alpha R +
\beta$.

\clearpage

\begin{figure}
\centering
\includegraphics[angle=0.0,width=15cm]{potential}
\caption{} \label{potential}
\end{figure}


\begin{thebibliography}{99}
\bibitem{Goldstone81} J. Goldstone and F. Wilczek, Phys. Rev. Lett.
\textbf{47}, 986 (1981).

\bibitem{Nussinov07} Z. Nussinov, C.D. Batista, B. Normand, and S.A. Trugman,
Phys. Rev. B \textbf{75}, 094411 (2007).

\bibitem{Baskaran03} G. Baskaran, Phys. Rev. B \textbf{68}, 212409 (2003).

\bibitem{Antonio07} A.R. Moura, A.R. Pereira, and A.S.T. Pires, Phys. Rev. B \textbf{75},
014431 (2007).

\bibitem{Castelnovo08} C. Castelnovo, R. Moessner, and  S. L. Sondhi, Nature
\textbf{451}, 42 (2008).

\bibitem{Joubert09} L.D.C. Joubert and P.C.W. Holdsworth, Nature
Phys. \textbf{5}, 258 (2009).

\bibitem{Wang06} R.F. Wang, C. Nisoli, R.S. Freitas, J. Li, W. McConville, B.J. Cooley,
M.S. Lund, N. Samarth, C. Leighton, V.H. Crespi, and P. Schiffer,
Nature \textbf{439}, 303 (2006).

\bibitem{Remhof08} A. Remhof, A. Shumann, A. Westphalen, H. Zabel, N. Mikuszeit,
E.Y. Vedmedenko, T. Last, and T. Kunze, Phys. Rev. B \textbf{77},
134409 (2008).

\bibitem{Ke08} X. Ke, J. Le, C. Nisoli, P.E. Lammert, W. McConville, R.F. Wang,
V.H. Crespi, and P. Schiffer, Phys. Rev. Lett. \textbf{101}, 037205
(2008).

\bibitem{Moller06} G. M\"{o}ller and R. Moessner, Phys. Rev. Lett.
\textbf{96}, 237202 (2006).

\bibitem{Weis03} J.-J. Weis, J. Phys.: Cond. Matt. \textbf{15}, S1471 (2003).

\bibitem{Wang01} Z. Wang and C. Holm, The J. Chem. Phys. \textbf{115}, 6351 (2001).

\bibitem{Landau05} D. Landau and  K. Binder, A guide to Monte Carlo
Simulations in Statistical Physics, Cabridge University Press, New
York (2005).

\bibitem{Tobochnik79} J. Tobochnik and G.V. Chester, Phys. Rev. B
\textbf{20}, 3761 (1979).

\bibitem{Kondev98} J.L. Jacobsen and J. Kondev, Nucl. Phys. B \textbf{532},
635 (1998).

\bibitem{Fisher63} M.F. Fisher and J. Stepheson, Phys. Rev.
\textbf{132}, 1411 (1963).

\bibitem{Moore99} C. Moore, M.G. Nordahl, N. Minar, and R. Shalizi,
Phys. Rev. E \textbf{60}, 5344 (1999).

\bibitem{Henley05} C.L. Henley, Phys. Rev. B \textbf{71}, 014424
(2005).

\bibitem{Teitelboim} M. Henneaux and C. Teitelboim, Phys. Rev. Lett. \textbf{56},
689 (1986).

\bibitem{Pisarski} R. Pisarski, Phys. Rev. D \textbf{34}, 3851 (1986).

\bibitem{PRD2001} W.A. Moura-Melo and J.A. Helay\"el-Neto, Phys. Rev.
D \textbf{63}, 065013 (2001).
\end{thebibliography}
\end{document}